\documentclass[preprint,aps]{revtex4}
\usepackage{amsmath}
\usepackage[dvips,xdvi]{graphicx}
\usepackage{amssymb}
\usepackage{epsfig}

\begin{document}
\title{Electric-field control of exciton fine structure: atomic scale manipulation of exchange
splitting
}

\author{Garnett W. Bryant$^1$}
\email{garnett.bryant@nist.gov}
\author{Natalia Malkova$^1$}
\author{James Sims$^2$}

\affiliation{$^1$Quantum Measurement Division and Joint Quantum Institute, 
National Institute of Standards and Technology, 100 Bureau Drive, Gaithersburg, 
Maryland 20899-8423\\
$^2$Information Technology Laboratory, National Institute of Standards 
and Technology, 100 Bureau Drive, Gaithersburg, Maryland 20899-8911\\
}
\date{\today}

\begin{abstract}
We use atomistic tight-binding theory with a configuration interaction description of Coulomb and exchange effects to describe excitons in symmetric quantum dots in a vertical electric field. We show that field-induced manipulation of exciton orientation and phase produces a drastic reduction of fine structure splitting, an anticrossing, and a 90 degree rotation of polarization, similar to experiment. An {\it atomistic} analysis is needed to explain how exciton reorientation modifies anisotropic exchange and fine structure splitting without significantly altering other splittings.

\end{abstract}

\maketitle

Tremendous effort has been made to control excitons in self-asembled semiconductor quantum dots (QD) using vertical electric~\cite{bennett10, boyer10, trotta12, luo12} and in-plane electric~\cite{kowalik05, gerardot07, reimer08, mar10}, magnetic~\cite{stevenson06}, optical~\cite{muller08, muller09}, and strain~\cite{trotta12, nakaoka03, nakaoka04, gell08, seidl06, ding10, plumhof11, singh10, bryant10, bryant11, gong11, wang12} fields, annealing~\cite{young05} and crystal symmetry~\cite{singh09, germanis12} to manipulate exciton fine-structure splitting and polarization.  Such control enables entangled photon generation by biexciton cascade~\cite{benson00}, coherent state manipulation, and transfer between "flying" photonic qubits and "stationary" solid-state qubits needed for quantum information processing. 

Experiments have demonstrated that vertical electric fields can coherently control exciton states in QDs, revealing a sharp, field-induced anticrossing of the ground-state, bright excitons with a large reduction of fine structure splitting and rotation of the polarization by 90 degrees at the anticrossing~\cite{bennett10, boyer10, trotta12}. Such experiments have successfully demonstrated control of the exciton but they do not reveal how that control is achieved, i.e. what about the exciton is being manipulated. Typically, it is thought that the exciton is modified by changing the asymmetry of the exciton that is imposed by lateral asymmetry of the QD or the underlying crystal lattice. However, an applied vertical field does not alter the lateral symmetry of the states. In this letter, we show that a vertical field modifies the vertical distribution of the exciton orientation and phase, atomic plane by atomic plane in the QD. This modification of exciton spatial structure leads to dramatic change in the anisotropic exchange which controls exciton fine structure and polarization.

\begin{figure}[htbp]
\includegraphics*[width=55mm]{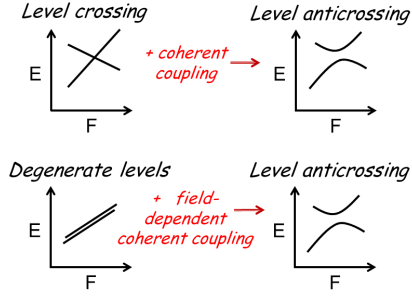}
\caption{Level anticrossing from two different mechanisms: (top) field-dependent level crossing plus coherent coupling to open the crossing 
and (bottom) two degenerate levels split by a field-dependent coupling with the minimum splitting determined by the minimum  
coupling. 
} \label{fig1} \end{figure}

The simplest model for understanding the exciton anticrossing~\cite{bennett10} is the traditional model of an anticrossing: two field-dependent levels, representing the two bright excitons, cross as the field is varied. A residual, field-independent coupling opens the crossing into an anticrossing when the levels are resonant (Fig.~\ref{fig1}). This two-level model arises~\cite{luo12} if the bright excitons are Stark-shifted by the field but there is no other change in exciton energies, i.e. the binding and exchange energies are unaffected by the field. There is a second model that can produce a similar anticrossing: two field-dependent, but degenerate levels are {\it split} by a {\it field-dependent} coupling. The mimimum coupling determines the position of the anticrossing (Fig.~\ref{fig1}). A straightforward analysis shows that there is no unitary transformation between these two models. They are different two-level models. In this letter, we show that the second model provides the {\it microscopic} understanding of the observed exciton control~\cite{bennett10}. Field-control of the exciton is realized by field-control of the anisotropic exchange coupling. Control of the exchange is essential and cannot be ignored. 

In typical QDs, the lowest electron and hole state are doubly degenerate due to spin.  As a result, the lowest electron-hole pair state is four-fold degenerate. Coulomb and exchange effects split the pair ground state into four exciton states. Two are dark excitons (DE), optically forbidden because they are made from the pair states with spin not conserved by excitation. We are concerned with the control of the two bright excitons (BE). The two lowest BE states are made primarily by mixing the two lowest degenerate  electron-hole pair states with total angular momentum, $J_z = \pm 1$, excited optically in a spin-conserving transition with circularly polarized light ($\Delta J_z = \pm 1$).  Mixing higher pair states is minimal. Using the two pair states that make up the BEs as a basis gives the following  two-level Hamiltonian for the BEs,
\[ H_{BE} = \left( \begin{array}{lr} E_{eh} + V_{coul} + V_{exch}^{1,1} & 
V_{exch}^{1,-1} \\ V_{exch}^{1,-1 \ast} & E_{eh} + V_{coul} + V_{exch}^{1,1}
 \end{array} \right).\]
$E_{eh}$ is the electron-hole pair energy, Stark shifted by the applied vertical field $F$.
The (real) coulomb energy $V_{coul}$ determines the exciton binding
energy, depends on $F$ but is independent of $J_z$ and does not mix states with different $J_z$.
The DE-BE splitting is determined by a (real) isotropic
exchange coupling, $V_{exch}^{1,1}$, also depends on $F$, is independent of $J_z$ and does not mix $J_z$. 
Consequently, the diagonal energies are field-dependent but degenerate. 
The exchange splitting between BEs is determined solely by the magnitude of the (complex) field-dependent
off-diagonal anisotropic exchange coupling, $V_{exch}^{1,-1}$, which mixes $J_z$. Here, we show that this model provides an understanding of the observed exciton control and explain the origin of the field-dependence of $V_{exch}^{1,-1}$.
More general models, with both non-degenerate field-dependent diagonal energies and field-dependent off-diagonal couplings have been proposed to describe exciton control~\cite{trotta12, gong11, wang12}. Unitary transformations can convert the model we consider with degenerate diagonal energies into the more general model by a partial diagonalization, so the models describe the same control. Fitting simple models to more complete calculations supports the models but does not explain the microscopic origin of exciton control.~\cite{trotta12, luo12, gong11, wang12} We show here that microscopic control of the exciton is achieved by atomic scale control of $V_{exch}^{1,-1}$ via control of the exciton orientation and phase.

We use atomistic tight-binding theory to describe electrons and holes in the QDs~\cite{bryant10, bryant11, diaz07, jaskolski06, diaz06} and a configuration interaction (CI) approach to describe excitons~\cite{bryant10, bryant11, leung97, franceschetti99, lee01} to establish the microscopic basis for the second model. The tight-binding model includes $sp^3s^*$ orbitals, nearest-neighbor coupling, spin-orbit effects and strain from lattice mismatch. Strain relaxation is included via atomistic valence force field theory. An atomistic model is needed to fully describe anisotropic exchange splitting~\cite{bester03, singh09, mlinar09}. While splitting arises if the QD has geometrical asymmetry~\cite{takagahara00}, splitting can arise, even for QDs with in-plane geometrical symmetry, because the atomic lattice breaks symmetry~\cite{bester03, singh09}. Atomic-scale resolution of electron, hole and exciton states will also be critical for explaining the field-dependence of the coupling. This is a striking example where {\it atomistic} effects play an essential role in determining the manipulation of {\it nanoscale} QD excitations. To identify the field-dependence, we study the InAs/GaAs, square-based, pyramidal QDs previously considered to investigate effects of mechanical strain on QDs~\cite{bryant10, bryant11}. The QDs are 
symmetric, so anisotropic exchange is induced by symmetry breaking by the lattice. 

\begin{figure}[htbp]
\includegraphics*[width=55mm]{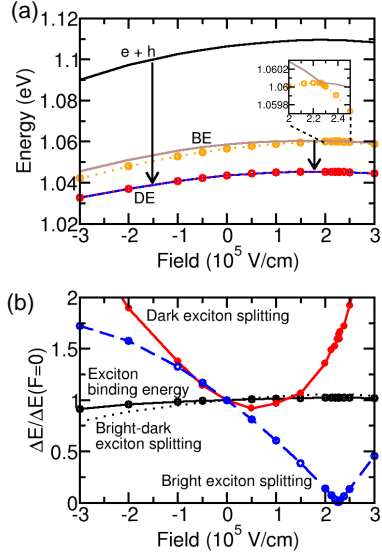}
\caption{(a)Field-dependence of the exciton energies: (e+h) Electon plus hole single-particle energy, (BE) the two lowest bright excitons, and (DE) the two, nearly-degenerate, dark excitons. The inset shows a blowup of the BE crossing. Exciton binding energy (long arrow) and DE-BE splitting (short arrow) are shown. (b) Energy splittings, $\Delta E$, normalized to the  zero-field splittings so all splittings fit on the same scale.
} \label{fig2} \end{figure}

The field dependence of the exciton levels, found from the full CI calcuation, is shown in Fig.~\ref{fig2}(a). Energies of the nearly degenerate bright and dark excitons follow the field-dependent Stark shifts of the lowest electron-hole pair level $E_{eh}$ with the exciton binding energy and BE-DE splitting indicated by arrows. Fig.~\ref{fig2}(b) shows that the binding energy and BE-DE splitting change slowly with field. However, the splittings that lift the degeneracy of the BE and of the DE show a much stronger dependence on $F$, each with a well-defined, deep minimum. The BE splitting, known as the fine-structure splitting, nearly vanishes for $F$ near 200 kV/cm. Near this resonance, the calculated field dependence is nearly linear, as seen experimentally,~\cite{bennett10} with a slope comparable to experiment. In addition, the linear polarization of the BEs (not shown) rotates by 90 degrees at resonance, as in the experiments.       

\begin{figure}[htbp]
\includegraphics*[width=55mm]{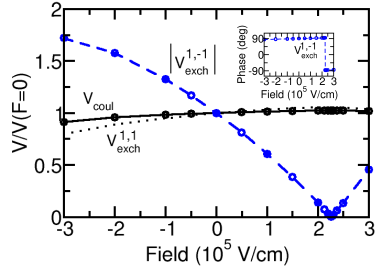}
\caption{Field-dependence of $V_{coul}$, $V^{1,1}_{exch}$ and the magnitude of $V^{1,-1}_{exch}$. The inset shows the phase of $V^{1,-1}_{exch}$.
} \label{fig3} \end{figure}

When the full CI calcuation is restricted to the two-level model, we extract $V_{coul}$, $V_{exch}^{1,1}$ and $V_{exch}^{1,-1}$, shown in Fig.~\ref{fig3}. $V_{coul}$ gives the binding energy, $V_{exch}^{1,1}$ the BE-DE splitting and $|V_{exch}^{1,-1}|$ the fine structure splitting. Fig.~\ref{fig3} shows that the two-level model describes well the exciton splittings. In the two-level model, the polarization of the exciton is determined by the phase of $V_{exch}^{1,-1}$. The absolute phase of $V_{exch}^{1,-1}$ is arbitrary. The phases of the two circularly-polarized ($J_z = \pm 1$) pair states used as BE basis states are chosen so that the pair-state polarizations are proportional to $x \pm \imath y$, where $x$ and $y$ are the directions of both the QD base and the crystal axes. With this choice of phases, excitons formed by coupling the two basis states have the expected linear polarization along the diagonals of the QD when $V_{exch}^{1,-1}$ is imaginary. As shown in Fig.~\ref{fig3}, $V_{exch}^{1,-1}$ is imaginary and changes sign when the fine structure splitting vanishes, leading to the observed 90 degree rotation of diagonal polarization. These results show that the simple model captures the essense of the exciton control and that control is determined by the field dependence of both the magnitude and phase of $V_{exch}^{1,-1}$.

\begin{figure}[htbp]
\includegraphics*[width=55mm]{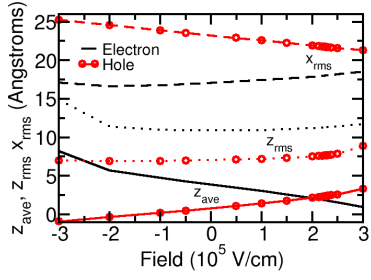}
\caption{Vertical position $z_{ave}$ and root-mean-square vertical $z_{rms}$ and lateral $x_{rms}$ spread of the lowest electron and hole states. The QD/wetting layer interface is at $z = 0$. 
} \label{fig4} \end{figure}

To appreciate how $F$ changes the couplings, recall~\cite{bryant10, bryant11} that $V_{coul}$ couples
electron density ($\rho_e({\textbf r}) = |\phi_e({\textbf r})|^2$, for electron state $\phi_e({\textbf r})$)
with hole density ($\rho_h({\textbf r^\prime}) = |\phi_h({\textbf r^\prime})|^2$):
\[V_{coul} = -\int \frac{\rho_e({\textbf r})\rho_h({\textbf r^\prime})}
{\epsilon({\textbf r},{\textbf r^\prime})|{\textbf r}-{\textbf r^\prime}|},\] 
with the local screening given by $\epsilon({\textbf r},{\textbf r^\prime})$. $V_{exch}^{1,1}$ couples polarization density  $P({\textbf r})=\phi_e({\textbf r})\phi_h({\textbf r})$ at ${\textbf r}$ with the \textit{conjugate} polarization density at ${\textbf r^\prime}$:
\[V_{exch}^{1,1} = \int \frac{P({\textbf r}) P^{\ast}({\textbf r^\prime})}
{\epsilon({\textbf r},{\textbf r^\prime})|{\textbf r}-{\textbf r^\prime}|}.\]
$V_{exch}^{1,1}$ depends mostly on the interaction weighted average of
$|P({\textbf r})|$ and $|P({\textbf r^\prime})|$ and weakly
on the phase difference between $P({\textbf r})$ and $P({\textbf r^\prime})$. Fig.~\ref{fig4} shows that the size of the electron and hole states changes little with $F$, consistent with the weak field-induced changes in $V_{coul}$ and  $V_{exch}^{1,1}$.
For $V_{exch}^{1,-1}$, the coupling is between
$P({\textbf r})$ and $P({\textbf r^\prime})$:
\[V_{exch}^{1,-1} = \int \frac{P({\textbf r}) P({\textbf r^\prime})}
{\epsilon({\textbf r},{\textbf r^\prime})|{\textbf r}-{\textbf r^\prime}|}.\]  
$V_{exch}^{1,-1}$ and $V_{exch}^{1,1}$ depend similarly on $|P({\textbf r})|$, so field-induced size changes of the states have little effect on the anisotropic exchange. However, $V_{exch}^{1,-1}$
depends on the sum of the phases of $P({\textbf r})$ and $P({\textbf r^\prime})$. Thus, field-induced changes in the phase of $P({\textbf r})$ must control the field-dependence of $V_{exch}^{1,-1}$ and determine its sign change. 

The most dramatic effect of the field on the electron and hole states is the shift of their vertical positions in the QD. The electron is pushed down toward the wetting layer and the hole up more into the QD as $F$ increases, see Fig.~\ref{fig4}. They cross at an $F$ close to the sign change of $V_{exch}^{1,-1}$. The close correspondence of these two crossings may be coincidental. Nonetheless, the phase control of $V_{exch}^{1,-1}$ must come from $F$ shifting the vertical position of the electron and hole in the QD. 

\begin{figure}[htbp]
\includegraphics*[width=55mm]{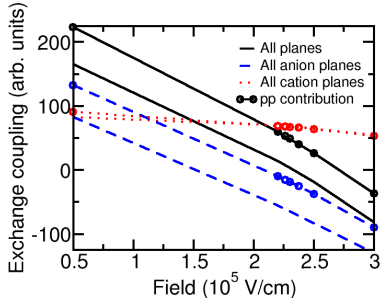}
\caption{Spatial and orbital contributions to $Im(V_{exch}^{1,-1})$: full calculation (all sites and orbitals), lateral cation planes, lateral anion planes. Curves with circles are the contributions that come from the $p$-orbitals in $P$ and $U_{exch}$.
} \label{fig5} \end{figure}

In the tight binding model, $V_{exch}^{1,-1}$ can be expressed as a sum over atomic sites $i$:  
\[V_{exch}^{1,-1} = \sum_i P({\textbf r_i}) U_{exch}({\textbf r_i}).\]
Ignoring the spatial dependence of the screening
\[U_{exch}({\textbf r_i}) = \sum_{j\neq i} \frac{P({\textbf r_j})}{\epsilon|{\textbf r_i}-{\textbf r_j}|}.\]
Here we have also ignored on-site contributions to $V_{exch}^{1,-1}$ because they are minor and change little with varying $F$. Terms that couple nearest neighbors and next-nearest neighbors are small, not changing sign for the $F$ considered. However, the remainining longer range contributions produce the field dependence seen in Fig.~\ref{fig3}. The contribution to $V_{exch}^{1,-1}$ from 
different atomic sites can be obtained by restricting the sum over $i$ appropriately. Fig.~\ref{fig5} compares the calculation for $V_{exch}^{1,-1}$ in this simplified form when including the contribution from all atomic sites with the contributions only from the lateral ($x,y$) anion planes and from the lateral cation planes. The contribution from the anion lateral planes parallels the full result. The sign change of $V_{exch}^{1,-1}$ occurs because the contribution from the anion planes changes sign. The nearly constant contribution of the cation planes shifts $V_{exch}^{1,-1}$ to more positive values, thereby moving the zero crossing to higher $F$. Both contributions are important, but each plays a different role in defining $V_{exch}^{1,-1}$. The orbital contributions can be isolated by restricting the sum that defines $P$ at each site to include only the contribution for those orbitals. Fig.~\ref{fig5} shows the $p$-orbital contribution to the exchange coupling that comes from $p$ orbitals in $P$ and $U_{exch}$. The $p$-orbital contribution also parallels the full result and defines the $F$ dependence. The other orbital contributions provide a shift that lowers the crossing $F$. These results show that atomic and orbital resolution is essential in determining $V_{exch}^{1,-1}$.  

The phase of $V_{exch}^{1,-1}$ is determined by summing $P({\textbf r_i}) U_{exch}({\textbf r_i})$ over all sites. The sum depends on the spatial dependence of the phase of $P({\textbf r_i})$ and $U_{exch}({\textbf r_i})$. The phase of $U_{exch}({\textbf r_i})$ is determined mostly by the phase of $P$ at sites near ${\textbf r_i}$. Heuristically, the phase of $V_{exch}^{1,-1}$ is then determined from the phase of $P^2({\textbf r_i})$. The phase of $P$ comes from the phase of the electron and hole states. The phase of the electron state, made primarily from $s$-orbitals on cation sites, should be independent of $\theta$, the angle of rotation around the vertical axis of the dot. The phase of the hole state, made primarily from $p$-orbitals on anion sites, should vary approximately as $\exp(\imath L \theta)$. For holes, the spatial angular momentum is $L = 1$ in the tight binding model (i.e. the phase resides in the coefficients of the orbital expansion because the orbitals are oriented the same on each site). Consequently, the phase of $P$ varies as $\exp(\imath \theta)$, the phase of $P U_{exch}$ as $\exp(\imath 2 \theta)$ and points that are rotated by $\theta = \pi/2$ should give contributions with the opposite sign. In particular, points near the $[1\bar{1}0]$ and $[110]$ diagonals should give contributions with similar magnitude but opposite sign. This is verified by explicit analysis of $P$ and $U_{exch}$. If the QD were symmetric and the crystal lattice did not break symmetry, then these contributions would have the same magnitude and would cancel. When lateral symmetry is broken, the contributions need not cancel and the sign of $V_{exch}^{1,-1}$ will depend on the orientation of $P$, how this orientation changes vertically in the dot, and how much the contributions of opposite sign cancel.

\begin{figure}[htbp]
\includegraphics*[width=55mm]{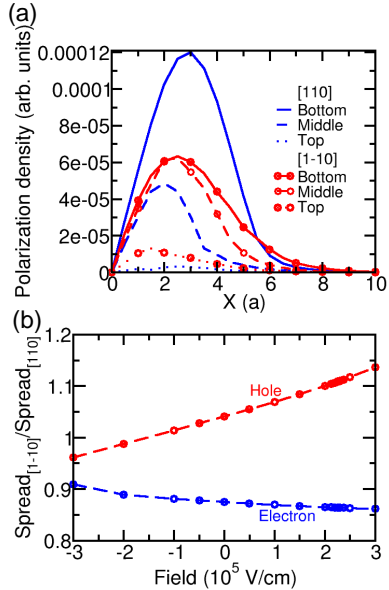}
\caption{(a) Spatial dependence of $P$ along $[110]$ and $[1\bar{1}0]$ for anion layers at the bottom, middle and top of the QD for $F = 50$ kV/cm. (b) Field dependence of the spread of the lowest electron and hole state along $[1\bar{1}0]$ compared to $[110]$. 
} \label{fig6} \end{figure}

The orientation of the hole and electron determines how $P$ is oriented. On anion layers, the hole is oriented along $[110]$ in the wetting layer
but along $[1\bar{1}0]$ inside the QD, more strongly so near the top of the QD. On cation layers, the hole is oriented along $[1\bar{1}0]$. The electron is always oriented along $[110]$ on anion layers, although less so at the top of the dot, and along $[110]$ on the cation layers. The change in orientation of the hole on anion layers leads to a rotation of $P$ in the QD, see Fig.~\ref{fig6}. Near the bottom of the QD, $P$ is oriented along $[110]$ on anion layers. Near the top of the QD, $P$ is oriented along $[1\bar{1}0]$ on anion layers. Contributions to $V_{exch}^{1,-1}$ from anion layers at the bottom and top of the dot have opposite signs. When $F$ is increased, the hole is pushed up more into the dot, the hole orientation rotates from $[110]$ to $[1\bar{1}0]$ (see Fig.~\ref{fig6}) and the contribution to $V_{exch}^{1,-1}$ from anion layers changes sign. This leads to the sign change of $V_{exch}^{1,-1}$. On cation layers, the orientation of $P$ changes much less. The  contributions to $V_{exch}^{1,-1}$ from cation layers have the same sign throughout most of the dot. As a result, the total cation contribution to $V_{exch}^{1,-1}$ changes slowly as $F$ increases and does not change sign for the range of $F$ considered here. This reorientation of the hole and the resulting change in the spatial distribution of the exciton phase, as revealed in $P$, determines the anisotropic exchange coupling and explains the $F$-dependent control of exciton fine structure splitting and polarization. 

In summary, atomistic theory has been used to show how applied vertical fields drastically reduces exciton fine structure splitting in QDs, with a narrow anticrossing and a 90 degree rotation of polarization. Control is achieved by field-dependent manipulation of the anisotropic exchange. The exchange is controlled by field-induced changes in the spatial distribution of the exciton orientation and phase that results from the rotation of the hole as it is pushed up into the QD by the field. This reorientation is an atomistic effect. The sign change in the anisotropic exchange arises from contribtions of anion layers. The contribution from cation layers is distinctly different, nearly constant with $F$, serving to shift the position of the crossing. Atomic layer sensitivity of QD fine structure is a challenge that must be overcome, but it also suggests new opportunity for atomic scale control of QD response. Here we have considered symmetric QDs, with the crystal lattice providing the symmetry breaking. Hole reorientation controls the sign of the exchange and nearly perfect cancellation of the exchange is possible. The same atomic-scale analysis for QDs with geometrical asymmetry has been left for further study.


\begin{thebibliography}{23}
\expandafter\ifx\csname
natexlab\endcsname\relax\def\natexlab#1{#1}\fi
\expandafter\ifx\csname bibnamefont\endcsname\relax
  \def\bibnamefont#1{#1}\fi
\expandafter\ifx\csname bibfnamefont\endcsname\relax
  \def\bibfnamefont#1{#1}\fi
\expandafter\ifx\csname citenamefont\endcsname\relax
  \def\citenamefont#1{#1}\fi
\expandafter\ifx\csname url\endcsname\relax
  \def\url#1{\texttt{#1}}\fi
\expandafter\ifx\csname
urlprefix\endcsname\relax\def\urlprefix{URL }\fi
\providecommand{\bibinfo}[2]{#2}
\providecommand{\eprint}[2][]{\url{#2}}

\bibitem{bennett10} A. J. Bennett, M. A. Pooley, R. M. Stevenson, M. B. Ward, R. B. Patel, 
A. Boyer de la Giroday, N. Sk\"{o}ld, I. Farrer, C. A. Nicoll, D. A. Ritchie, and A. J. Shields, Nature Phys. {\bf 6}, 947 (2010).

\bibitem{boyer10} A. Boyer de la Giroday, A. J. Bennett, M. A. Pooley, R. M. Stevenson, N. Sk\"{o}ld, R. B. Patel, I. Farrer,
D. A. Ritchie, and A. J. Shields, Phys. Rev. B {\bf 82}, 241301 (2010).

\bibitem{trotta12} R. Trotta, E. Zallo, C. Ortix, P. Atkinson, J. D. Plumhof, J. van den Brink, A. Rastelli, and O. G. Schmidt, Phys. Rev. Lett. {\bf 109}, 147401 (2012).

\bibitem{luo12} J.-W. Luo, R. Singh, A. Zunger, and G. Bester, Phys. Rev. B {\bf 86}, 161302 (2012).

\bibitem{kowalik05} K. Kowalik, O. Krebs, A. Lemaitre, S. Laurent, P. Senellart, P. Voisin, and J. A. Gaj, Appl. Phys. Lett. {\bf 86}, 041907 (2005).

\bibitem{gerardot07} B. G. Gerardot, S. Seidl, P. A. Dalgarno, R. J. Warburton, D. Granados, J. M. Garcia, K. Kowalik, and O. Krebs, Appl. Phys. Lett. {\bf 90}, 041101 (2007).

\bibitem{reimer08} M. E. Reimer, M. Korkusinski, D. Dalacu, J. Lefebvre, J. Lapointe, P. J. Poole, G. C. Aers, W. R. McKinnon, P. Hawrylak, and R. L. Williams, Phys. Rev. B {\bf 78}, 195301 (2008).

\bibitem{mar10} J. D. Mar, X. L. Xu, J. S. Sandhu, A. C. Irvine, M. Hopkinson, and D. A. Williams, Appl. Phys. Lett. {\bf 97}, 221108 (2010).

\bibitem{stevenson06} R. M. Stevenson, R. J. Young, P. See, D. G. Gevaux, K. Cooper, P. Atkinson, I. Farrer, D. A. Ritchie, and A. J. Shields, Phys. Rev. B {\bf 73}, 033306 (2006).

\bibitem{muller08} A. Muller, W. Fang, J. Lawall, and G. S. Solomon, Phys. Rev. Lett. {\bf 101}, 027401 (2008).

\bibitem{muller09} A. Muller, W. Fang, J. Lawall, and G. S. Solomon, Phys. Rev. Lett. {\bf 103}, 217402 (2009).

\bibitem{nakaoka03} T. Nakaoka, T. Kakitsuka, T. Saito, S. Kako, S. Ishida, M. Nisioka, Y. Yoshikuni, and Y. Arakawa, J. Appl. Phys. {\bf 94}, 6812 (2003).

\bibitem{nakaoka04} T. Nakaoka, T. Kakitsuka, T. Saito, and Y. Arakawa, Appl. Phys. Lett. {\bf 84}, 1392 (2004).

\bibitem{gell08} J. R. Gell, M. B. Ward, R. J. Young, R. M. Stevenson, P. Atkinson, D. Anderson, G. A. C. Jones, D. A. Ritchie, and A. J. Shields, Appl. Phys. Lett. {\bf 93}, 081115 (2008).

\bibitem{seidl06} S. Seidl, M. Kroner, A. Hogele, K. Karrai, R. J. Warburton, B. D. Gerardot, and P. M. Petroff, Appl. Phys. Lett. {\bf 88}, 203113 (2006).

\bibitem{ding10} F. Ding, R. Singh, J. D. Plumhof, T. Zander, V. K\v{r}\'{a}pek, Y. H. Chen, M. Benyoucef, V. Zwiller, K. D\"{o}rr,
G. Bester, A. Rastelli, and O. G. Schmidt, Phys. Rev. Lett. {\bf 104}, 067405 (2010).

\bibitem{plumhof11} J. D. Plumhof, V. K\v{r}\'{a}pek, F. Ding, K. D. J\"{o}ns, R. Hafenbrak, P. Klenovsk\'{y}, A. Herklotz,
K. D\"{o}rr, P. Michler, A. Rastelli, and O. G. Schmidt, Phys. Rev. B {\bf 83}, 121302 (2011).

\bibitem{singh10} R. Singh and G. Bester, Phys. Rev. Lett. {\bf 104}, 196803 (2010).

\bibitem{bryant10} G. W. Bryant, M. Zieli\'{n}ski, N. Malkova, J. Sims, W. Jask\'{o}lski, and J. Aizpurua, Phys. Rev. Lett. {\bf 105}, 067404 (2010).

\bibitem{bryant11} G. W. Bryant, M. Zieli\'{n}ski, N. Malkova, J. Sims, W. Jask\'{o}lski, and J. Aizpurua, Phys. Rev. B {\bf 84}, 235412 (2011).

\bibitem{gong11} M. Gong, W. Zhang, G.-C. Guo, and L. He, Phys. Rev. Lett. {\bf 106}, 227401 (2011).

\bibitem{wang12} J. Wang, M. Gong, G.-C. Guo, and L. He, Appl. Phys. Lett. {\bf 101}, 63114 (2012).

\bibitem{young05} R. J. Young, R. M. Stevenson, A. J. Shields, P. Atkinson, K. Cooper, D. A. Ritchie, K. M. Groom, A. I. Tartakovskii, and M. S. Skolnick,, Phys. Rev. B {\bf 72}, 113305 (2005).

\bibitem{singh09} R. Singh and G. Bester, Phys. Rev. Lett. {\bf 103}, 063601 (2009).

\bibitem{germanis12} S. Germanis,  A. Beveratos, G. E. Dialynas, G. Deligeorgis, P. G. Savvidis, Z. Hatzopoulos, and N. T. Pelekanos, Phys. Rev. B {\bf 86}, 035323 (2012).

\bibitem{benson00} O. Benson, C. Santori, M. Pelton, and Y. Yamamoto, Phys. Rev. Lett. {\bf 84}, 2513 (2000).

\bibitem{diaz07} J. G. Diaz, G. W. Bryant, W. Jask\'{o}lski, and M. Zieli\'{n}ski, Phys. Rev. B {\bf 75}, 245433 (2007). 

\bibitem{jaskolski06} W. Jask\'olski, M. Zieli\'{n}ski, G. W. Bryant, and J. Aizpurua, Phys. Rev. B {\bf 74}, 195339 (2006).

\bibitem{diaz06} J. G. Diaz, M. Zieli\'{n}ski, W. Jask\'{o}lski, and G. W. Bryant, Phys. Rev. B {\bf 74}, 205309 (2006).

\bibitem{leung97} K. Leung and K. B. Whaley, Phys. Rev. B {\bf 56}, 7455 (1997).

\bibitem{franceschetti99} A. Franceschetti, H. Fu, L. W. Wang, and A. Zunger, Phys. Rev. B {\bf 60}, 1819 (1999).

\bibitem{lee01} S. Lee, L. J\"{o}nsson, J. W. Wilkins, G. W. Bryant, and G. Klimeck, Phys. Rev. B {\bf 63}, 195318 (2001).

\bibitem{bester03} G. Bester, S. Nair, and A. Zunger, Phys. Rev. B {\bf 67}, 161306 (2003).

\bibitem{mlinar09} V. Mlinar and A. Zunger, Phys. Rev. B {\bf 79}, 115416 (2009).

\bibitem{takagahara00} T. Takagahara, Phys. Rev. B {\bf 62}, 16840 (2000).

\end{thebibliography}
\end{document}